\documentclass[aps,prb,reprint,superscriptaddress]{revtex4-1}

\usepackage{graphicx}
\usepackage{amsmath}
\usepackage{textcomp} 
\usepackage{xcolor}
\usepackage{ulem}
\usepackage{amssymb}

\newcommand{\YIG}{Y$_\mathrm{3}$Fe$_\mathrm{5}$O$_\mathrm{12}$}
\newcommand{\g}{$g_\mathrm{eff}^{\uparrow \downarrow}$}

\begin{document}

\title{Large spin-wave bullet in a ferrimagnetic insulator driven by spin Hall effect}


\author{M.~B.~Jungfleisch}
\email{jungfleisch@anl.gov}
\affiliation{Materials Science Division, Argonne National Laboratory, Argonne IL 60439, USA}

\author{W.~Zhang}
\affiliation{Materials Science Division, Argonne National Laboratory, Argonne IL 60439, USA}

\author{J.~Sklenar}
\affiliation{Materials Science Division, Argonne National Laboratory, Argonne IL 60439, USA}
\affiliation{Department of Physics and Astronomy, Northwestern University, Evanston IL 60208, USA}

\author{J.~Ding}
\affiliation{Materials Science Division, Argonne National Laboratory, Argonne IL 60439, USA}

\author{W.~Jiang}
\affiliation{Materials Science Division, Argonne National Laboratory, Argonne IL 60439, USA}

\author{H.~Chang}
\affiliation{Department of Physics, Colorado State University, Fort Collins CO 80523, USA}

\author{F.~Y.~Fradin}
\affiliation{Materials Science Division, Argonne National Laboratory, Argonne IL 60439, USA}

\author{J.~E.~Pearson}
\affiliation{Materials Science Division, Argonne National Laboratory, Argonne IL 60439, USA}

 \author{J.~B.~Ketterson}
\affiliation{Department of Physics and Astronomy, Northwestern University, Evanston IL 60208, USA}

\author{V.~Novosad}
\affiliation{Materials Science Division, Argonne National Laboratory, Argonne IL 60439, USA}

\author{M.~Wu}
\affiliation{Department of Physics, Colorado State University, Fort Collins CO 80523, USA}

\author{A.~Hoffmann}
\affiliation{Materials Science Division, Argonne National Laboratory, Argonne IL 60439, USA}

\begin{abstract}
Due to its transverse nature, spin Hall effects (SHE) provide the possibility to excite and detect spin currents and  magnetization dynamics even in magnetic insulators. 
Magnetic insulators are outstanding materials for the investigation of nonlinear phenomena and for 
novel low power spintronics applications because of their extremely low Gilbert damping. 
Here, we report on the direct imaging of electrically driven spin-torque ferromagnetic resonance (ST-FMR) 
 in the ferrimagnetic insulator Y$_3$Fe$_5$O$_{12}$ based on the excitation and detection by SHEs. 
The driven spin dynamics in Y$_3$Fe$_5$O$_{12}$ is directly imaged by spatially-resolved microfocused Brillouin light scattering (BLS) spectroscopy. 
Previously, ST-FMR experiments assumed a uniform precession across the sample, which is not valid in our measurements. 
A strong spin-wave localization in the center of the sample is observed indicating the formation of a nonlinear, self-localized spin-wave `bullet'.


\end{abstract}
\date{\today}
\maketitle

Magnetic memory and logic devices rely on the efficient manipulation of the orientation of their magnetization using low power \cite{STT, Mesoscale_mag}. 
Recently, there has been revitalized interest in the ferrimagnetic insulator yttrium iron garnet (YIG, \YIG) motivated by the discovery of spintronic effects by combining this material and heavy metals such as Pt \cite{Kajiwara,Uchida,Jungfleisch_SSE,Jungfleisch,Jungfleisch_prop}. Its extremely small magnetic damping enables low power data transmission and processing on the basis of magnons, the elementary quanta of magnetic excitations. \cite{Hoffmann_book,Melkov_book, Kajiwara, Jungfleisch, Jungfleisch_prop, Pirro,Klein_PRL,Klein_APL}. In addition the low damping YIG also enables nonlinear phenomena where the superposition principle breaks down 
\cite{Melkov_book}. Previous work reported on the formation of spin-wave caustics \cite{Schneider}, Bose Einstein condensation of magnons \cite{BEC} and nonlinear mode conversion \cite{Kostylev_PRB} to name only a few. 
Recently, it has become possible to grow nanometer-thick YIG films, which allow the preparation of micro- and nanostructured devices \cite{Jungfleisch_prop, Pirro, Houchen, Klein_PRL, Klein_APL}. Therefore, the study of nonlinear spin dynamics in miniaturized YIG systems has only just begun.

Independent of the progress of the YIG film growth, the development in employing spin-orbit interaction in heavy metals \cite{Hoffmann,Zhang_JAP} and their alloys \cite{Zhang_AF} in contact with a ferromagnet (FM) has flourished. The SHE \cite{Hirsch, Dyakonov} can be used for the generation of strong current-driven torques on the magnetization in the FM layer. 
The resultant spin current can drive spin-torque ferromagnetic resonance (ST-FMR) in bilayers consisting of ferromagnetic and nonmagnetic metals and be detected by a homdyne mixing of the microwave signal with the anisotropic magnetoresistance \cite{Liu_ST-FMR}. 
Recent theories propose that ST-FMR can be extended to insulating FM/normal metal bilayers. Here, the detection of magnetization precession occurs by spin pumping and a rectification of the spin Hall magnetoresistance \cite{Chiba,Chiba2}. We showed recently that this rectification process is indeed possible in YIG/Pt bilayers \cite{Joe_PRL}. All previous analysis of electric measurements assume uniform precession across the sample \cite{Liu_ST-FMR,Mellnik_ST-FMR}. In order to validate this assumption it is highly desirable to image \textit{ac} current-driven spin dynamics spatially-resolved and frequency-resolved. These investigations provide interesting insights in the underlying physics, such as whether bulk or edge modes are preferably excited by ST-FMR or nonlinear spin dynamics may occur. 

\begin{figure}[b]
\includegraphics[width=1\columnwidth]{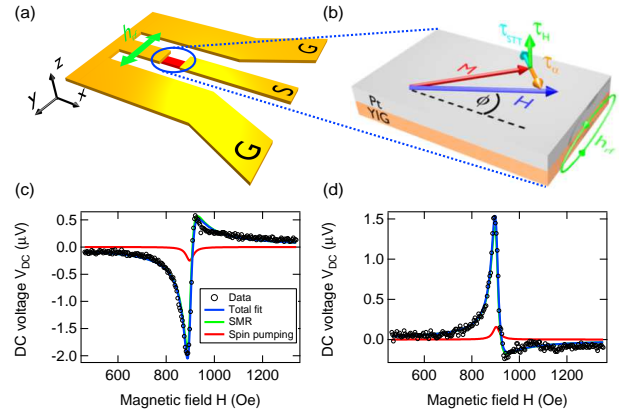}
\caption{\label{fig1} (a) Schematic of the ST-FMR experimental setup (b) ST-FMR mechanism in the YIG/Pt bilayer. The alternating \textit{rf} current drives an Oersted field $h_\mathrm{rf}$ exerting a field-like torque $\tau_\mathrm{H}$ on the magnetization $M$. At the same time a oscillatory transverse spin accumulation at the YIG/Pt interface is generated by the SHE which results in a damping-like torque $\tau_\mathrm{STT}$. (c) and (d) Typical \textit{dc} voltage spectra recorded at in-plane angles of $\phi = 30^\circ$ and $\phi = 240^\circ$ and $P= +10$~dBm.} 
\end{figure}

In this letter, we show experimentally the excitation of spin dynamics in microstructured magnetic insulators by the SHE of an adjacent heavy metal and observe the formation of a nonlinear, self-localized spin-wave intensity in the center of the sample \cite{Demidov_Nat12,Slavin_bullet,Urazhdin}. The magnetization dynamics in a nanometer-thick YIG layer is driven simultaneously by the Oersted field and a spin torque originating from a spin current generated by the SHE of an attached Pt layer. The dynamics is detected in two complementary ways: (1) Electrically, by a rectification mechanism of the spin Hall magnetoresistance (SMR) \cite{Nakayama,Schreier_STT,spin-diode} as well as by spin pumping \cite{Tserkovnyak, Azevedo_ISHE, Jungfleisch, Jungfleisch_prop, Kajiwara,Mosendz} and (2) Optically, by spatially-resolved Brillouin light scattering (BLS) microscopy \cite{BLS1}. The experimental findings are further validated by micromagnetic simulations \cite{mumax3}.

\begin{figure}[t]
\includegraphics[width=1.0\columnwidth]{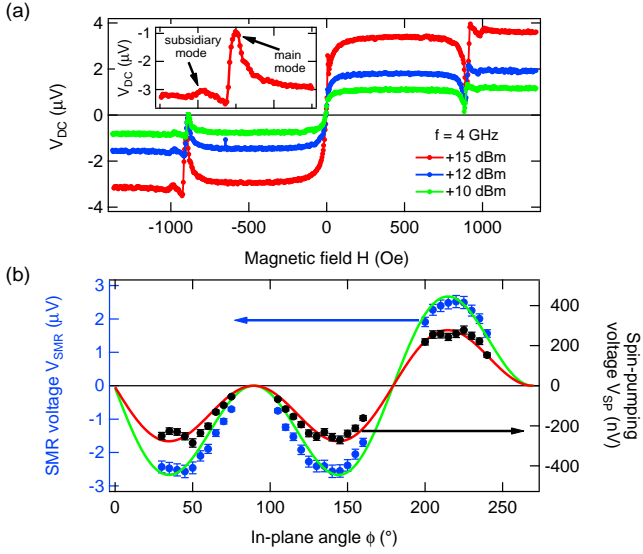}
\caption{\label{fig2} (a) Typical $V_\mathrm{DC}$ spectra at a constant frequency $f=4$~GHz for various applied microwave powers. The inset shows the resonance peak at $P=+15$~dBm. Two modes are detected. (b) In-plane angular dependence of the SMR, $V_\mathrm{SMR}$, and of the spin-pumping contribution, $V_\mathrm{SP}$, to the \textit{dc} voltage. The solid lines represent fits $\propto$ cos$~\phi~$sin~$2\phi$.} 
\end{figure}

YIG(40 nm)/Pt bilayers were fabricated by in-situ magnetron sputtering under high-purity argon atmosphere on single crystal gadolinium gallium garnet (GGG, Gd$_3$Ga$_5$O$_{12}$) substrates of 500 $\mu$m thickness with (111) orientation \cite{Houchen}. For the electrical measurements a Pt thickness of 2 nm was used, while for the optical investigations the thickness was 5 nm in order to minimize the influence of additional heating effects by the laser. In a subsequent fabrication process, stripes in the shape of $30 \times 5$ $\mu$m$^2$ (electrical measurements) and $5 \times 5$ $\mu$m$^2$ (optical measurements) were patterned by photolithography and ion milling \cite{Jungfleisch_prop}. A coplanar waveguide (CPW) made of Ti/Au (3 nm/120 nm) was structured on top of the bar allowing the signal line to serve as a lead for the YIG/Pt bar as illustrated in Fig.~\ref{fig1}(a). In this ST-FMR configuration a bias-T is utilized to allow for simultaneous transmission of a microwave signal with \textit{dc} voltage detection via lock-in technique across the Pt. For this purpose the amplitude of the \textit{rf} current is modulated at 3 kHz. We use a BLS microscope with a spatial resolution of 250 nm, where the laser spot is focused onto the sample and the frequency shift of the back reflected light is analyzed by a multi-pass tandem Fabry P\'erot interferometer \cite{BLS1}. 
The detected BLS intensity is proportional to the square of the dynamic magnetization, i.e., the spin-wave intensity.

In order to excite a dynamic response by ST-FMR in the YIG system a \textit{rf} signal is passed through the Pt layer. The magnetization dynamics is governed by a modified Landau-Lifshitz-Gilbert equation \cite{Chiba,Chiba2}:
\begin{equation}
\label{mod_LLG}
\frac{d\mathbf{M}}{dt}= - \vert\gamma\vert \mathbf{M}\times \mathbf{H}_\mathrm{eff}+\alpha \mathbf{M}\times \frac{d\mathbf{M}}{dt}+\frac{\vert\gamma\vert \hbar}{2e M_\mathrm{s}d_\mathrm{F}}\mathbf{J}_\mathrm{s},
\end{equation}
where $\gamma$ 
is the gyromagnetic ratio, $\mathbf{H}_\mathrm{eff}=\mathbf{h}_\mathrm{rf}+\mathbf{H}_\mathrm{D}+\mathbf{H}_\mathrm{}$ is the effective magnetic field including the microwave magnetic field $\mathbf{h}_\mathrm{rf}$, demagnetization fields $\mathbf{H}_\mathrm{D}$, and the bias magnetic field $\mathbf{H}$. $\alpha$ is the Gilbert damping parameter [the second term describes the damping torque $\tau_\mathrm{\alpha}$, Fig.~\ref{fig1}(b)] and $\mathbf{J}_\mathrm{s}$ is a transverse spin current at the interface generated by the SHE from the alternating charge current in the Pt layer \cite{Chiba,Chiba2}:
\begin{equation}
\label{J_s}
\mathbf{J}_\mathrm{s}=\frac{\mathrm{Re}(g_\mathrm{eff}^{\uparrow \downarrow})}{e}\mathbf{M}\times(\mathbf{M}\times\mu_\mathrm{s})+\frac{\mathrm{Im}(g_\mathrm{eff}^{\uparrow \downarrow})}{e}\mathbf{M}\times \mu_\mathrm{s}.
\end{equation}
Here, $g_\mathrm{eff}^{\uparrow \downarrow}$ is the effective spin-mixing conductance and $\mu_\mathrm{s}$ is the spin accumulation at the YIG/Pt interface. The first term in Eq.~(\ref{J_s}) describes an anti-damping-like torque $\tau_\mathrm{STT}$ and the second term is a field-like torque $\tau_\mathrm{H}$. As illustrated in Fig.~\ref{fig1}(b) and described by Eq.~(\ref{mod_LLG}) the magnetization is driven by the independent torque terms containing $h_\mathrm{rf}$ and $J_\mathrm{s}$.

\begin{figure}[b]
\includegraphics[width=1.0\columnwidth]{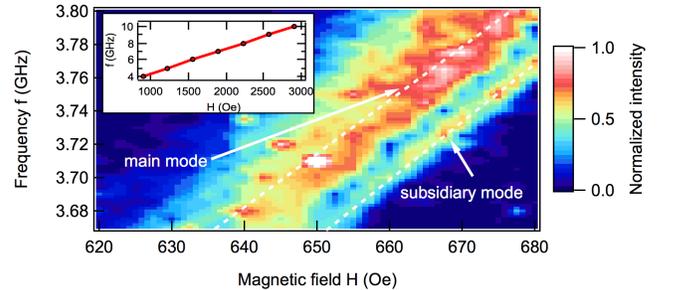}
\caption{\label{fig3} Color-coded dispersion relation measured by BLS microscopy. The laser spot was focused onto the center of the sample while the \textit{rf} frequency as well as magnetic field were varied. As for the electrical measurements two modes are detected by BLS. The inset shows corresponding field dependence of the resonance measured by electrical means.} 
\end{figure}

First, we describe the electrical characterization of the YIG/Pt bars by means of ST-FMR. Figure~\ref{fig1} (c) and (d) illustrate typical \textit{dc} voltage spectra; exemplarily, we show spectra recorded at in-plane angles of $\phi = 30^\circ$ and $\phi = 240^\circ$, with applied \textit{rf} power $P= +10$~dBm.  A signal is observed when the system is driven resonantly. The data is analyzed using the model proposed by Chiba et al. (supplementary information) \cite{Chiba,Chiba2}. According to the model, two signals contribute to the \textit{dc} voltage: (1) Spin pumping which manifests in a symmetric contribution to the Lorentzian lineshape. (2) Spin Hall magnetoresistance which is a superimposed symmetric and antisymmetric Lorentzian curve [Fig.~\ref{fig1}(c,d)]. 

Figure~\ref{fig2}(a) illustrates \textit{dc} voltage spectra at a fixed microwave frequency $f=4$~GHz for three different applied powers. The offset is due to the longitudinal spin Seebeck effect \cite{Uchida,Jungfleisch_SSE}  (see supplementary information) 
and does not affect the conclusions drawn from the resonance signal \cite{Jungfleisch_SSE}. The inset in Fig.~\ref{fig2}(a) shows the resonance peak at $P= +15$~dBm. Clearly, a less intense, secondary mode in addition to the main mode is detected. 
According to the Chiba model \cite{Chiba,Chiba2} the \textit{dc} voltage signal can be deconvoluted into a spin-pumping and a SMR contribution as also shown in Fig.~\ref{fig1}(c) and (d). To analyze the data employing the model we use a spin-mixing conductance of \g$ = 3.36 \times 10^{14}~\Omega^{-1}$ m$^{-2}$ and a spin-Hall angle of $\gamma_\mathrm{SHE} = 0.09$ \cite{Zhang_APL13}. A fit to the angular-dependent data yields a phase difference between Oersted field and the \textit{ac} current of $\delta = 64\pm 5^{\circ}$ [see Fig.~\ref{fig1}(c,d)]. 
Figure~\ref{fig2}(b) shows the angular dependences of the fitted spin-pumping and the SMR signals. The model predicts the same angular dependent behavior $\propto\mathrm{cos}\phi~\mathrm{sin}2\phi$ for spin pumping and SMR. As seen in Fig.~\ref{fig2}(b), we find a good agreement between theory (solid lines) and experiment for both curves. Please note that we observe a small, non-vanishing voltage at angles $\phi = n\cdot 90^\circ, n \in \mathbb{N}$, where the model suggests zero voltage \cite{Chiba,Chiba2,Joe_PRL}. In this angular range the model breaks down and the experimental data cannot be fitted (see supplementary information).


\begin{figure}[t]
\includegraphics[width=0.75\columnwidth]{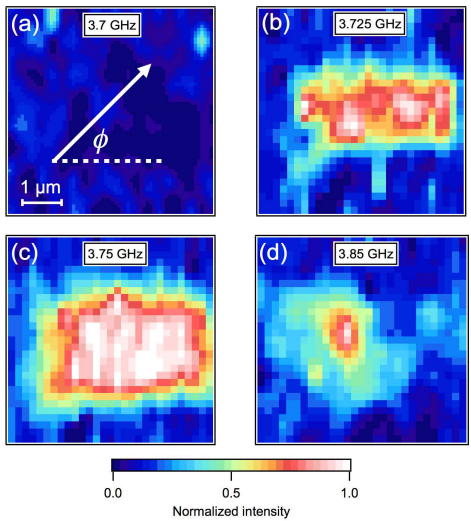}
\caption{\label{fig4} Spatially-resolved BLS map of the $5\times 5$ $\mu$m$^2$ large YIG/Pt sample. The magnetic field $H = 665$ Oe is applied at $\phi \sim 45^\circ$. (a) - (d) Driving microwave frequency increases from 3.7 GHz to 3.85 GHz, microwave power $P=+17$~dBm.} 
\end{figure}

In the following we compare the electrical measurements with the results obtained by BLS imaging. The optical measurements were performed on YIG(40 nm)/Pt(5 nm) bars having a lateral size of $5\times 5$~$\mu$m$^2$. The external magnetic field is applied at an angle of  $\phi \sim45^\circ$ where the \textit{dc} voltage detection is maximized [Fig.~\ref{fig2}(b)]. Figure~\ref{fig3} shows the dispersion relation measured by BLS in a false color-coded image where red indicates a high spin-wave intensity and the blue area shows the absence of spin waves. The measured dispersion is in agreement with the electrical measurements as shown in the inset: As the field increases the resonance shifts to higher frequencies as is expected from the Kittel equation, ${f= \frac{\vert\gamma\vert}{2\pi} \sqrt{H_\mathrm{}(H_\mathrm{}+4\pi M_\mathrm{eff})}}$,
where $M_\mathrm{eff}$ is the effective magnetization. 

As is apparent from Fig.~\ref{fig2}(a) and Fig.~\ref{fig3} magnetization dynamics can be excited in a certain bandwidth around the resonance which is determined by the specific device characteristics. Furthermore, both figures (electrical and optical detection) suggest that there is an additional mode below the main mode. At first, one might identify this mode as an edge mode \cite{edge_mode,Neusser}. However, this is not the case as it will be discussed below.

In order to spatially map the spin-wave intensity, the applied magnetic field is kept fixed at $H= 665$ Oe. Figure~\ref{fig4} illustrates the experimental observations in false color-coded images. At an excitation frequency below the resonance frequency, e.g., $f=3.7$~GHz no magnetization dynamics is detected [Fig.~\ref{fig4}(a)]. As the frequency increases the system is driven resonantly and a strong spin-wave intensity is observed from $f=3.725$~GHz to $f=3.8$~GHz, Fig.~\ref{fig4}(b,c). Increasing the frequency even further results in a diminished signal, Fig.~\ref{fig4}(d) for $f=3.85$~GHz. At even larger frequencies no magnetization dynamics is detected as it is also apparent from the dispersion illustrated in Fig.~\ref{fig3}. In conventional electrical ST-FMR measurements, a uniform spin-wave intensity distribution across the lateral sample dimensions is assumed. However, as our experimental results show, this assumption is not fulfilled: A strong spin-wave signal is localized in the center of the YIG/Pt bar. 
It is desirable to experimentally investigate at what minimum excitation power the formation of the localization occurs. However, in the investigated range of powers we always observe a localization in the center of the sample (see supplementary information). For \textit{rf} powers of less than +11 dBm the signal is below our noise-floor. 

\begin{figure}
\includegraphics[width=0.9\columnwidth]{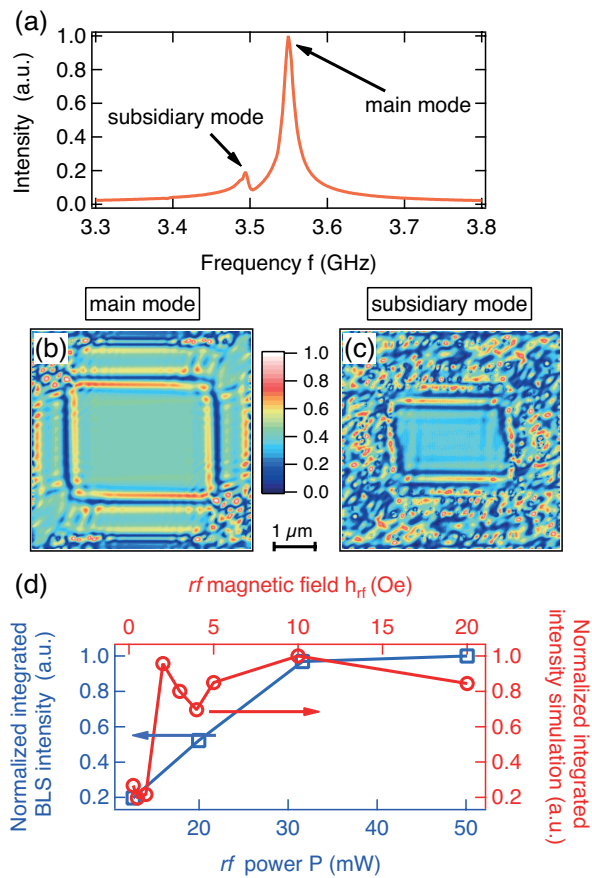}
\caption{\label{fig5} Micromagnetic simulations: (a) The spectrum reveals two modes. Spatially-resolved magnetization distribution of the main mode, (b), and the less intense, subsidiary mode, (c). (d) The normalized integrated BLS intensity saturates at high excitation powers $P$, which is validated by micromagnetic simulations at large driving \textit{rf} magnetic fields $h_\mathrm{rf}$.} 
\end{figure}

In spite of this experimental limitation, we also carried out micromagnetic simulations in order to gain further insight into the underlying magnetization dynamics. The simulations confirmed qualitatively the experimental observations as is depicted in Fig.~\ref{fig5}: Two modes can be identified in the simulations, Fig.~\ref{fig5}(a). In the low power regime, which is not accessible experimentally, we find that the spatial magnetization distribution of the main mode is almost uniform and the less intense subsidiary mode is localized at the edges ($h_\mathrm{rf}=0.25$~Oe, not shown). With increasing \textit{rf} power, the spatial distributions of both modes transform and at a threshold of $h_\mathrm{rf} \approx 1$~Oe a localization of both modes in the center of the sample is observed. Figure~\ref{fig5}(b,c) show the corresponding spatial dynamic magnetization distributions at $h_\mathrm{rf} = 5$~Oe and agree well with the experimental findings, Fig.~\ref{fig4}. 

This spatial profile can be understood as the formation of a nonlinear, self-localized `\textit{bullet}'-like spin-wave intensity caused by nonlinear cross coupling between eigenmodes in the system \cite{Kostylev_PRB}. This process is mainly determined by nonlinear spin-wave damping which transfers energy from the initially excited ferromagnetic resonance into other spin-wave modes rather than into the crystalline lattice \cite{Kostylev_PRB}. 
 To check this assumption, we plotted in Fig.~\ref{fig5}(d) the normalized integrated BLS intensity as well as the integrated spatial magnetization distribution as a function of the applied microwave power and the \textit{rf} magnetic field, respectively. Both integrated signals demonstrate a nonlinear behavior and saturate at high powers/microwave magnetic fields. This observation is a direct manifestation of nonlinear damping: energy is absorbed by the ferromagnetic resonance and redistributed to secondary spin-wave modes more and more effectively \cite{Kostylev_PRB}.

Until now, ST-FMR experiments assumed a uniform magnetization precession 
\cite{Liu_ST-FMR,Mellnik_ST-FMR,Chiba,Chiba2}. However, as our spatially-resolved BLS results demonstrate and confirmed by micromagnetic simulations, the driven lateral spin-wave intensity distribution in insulating FMs deviates from this simple model at higher excitation powers which are common in ST-FMR measurements. The formation of a localized spin-wave mode was not considered in previous ST-FMR experiments neither in metals nor in insulators. Our findings have direct consequences on the analysis and interpretation of ST-FMR experiments. The precession amplitude is not uniform across the sample implying that the effective spin-mixing conductance \g~ is actually an average over the sample cross section. In areas where the precession amplitude is large, \g~ is underestimated, whereas it is over estimated in low-intensity areas. This also complicates the determination of the spin-Hall angle from ST-FMR measurements.

Micromagnetic simulations show phase inhomogeneity, specifically around the perimeter of the mode.  The phase inhomogeneity tends to equally lag and lead the main uniform phase of the center mode; effectively the phase inhomogeneity then leads to no significant change to the lineshape. However, assuming the phase at the perimeter to be uniformly leading the bulk phase results in a correction to the lineshape that is still negligible because the effective area and amplitude where the phase is deviating is significantly smaller than the bulk area. Nevertheless, in general the issue of inhomogeneous phase distribution may complicate the analysis of electrical ST-FMR spectra, especially in smaller samples.


In conclusion, we demonstrated that the concept of ST-FMR can be extended to magnetic insulators where the formation of a nonlinear, self-localized spin-wave intensity driven by an \textit{ac} current was observed. We adopted an electrically-driven ST-FMR excitation and detection scheme in magnetic insulator (YIG)/heavy normal metal (Pt) bilayers that was originally developed for all-metallic systems. A \textit{dc} voltage in YIG/Pt bilayers was observed under resonance condition by a SMR-mediated spin-torque diode effect in agreement with theoretical predictions. Spatially-resolved BLS microscopy revealed a strong `\textit{bullet}'-like spin-wave localization in the center of the sample due to nonlinear cross coupling of eigenmodes in the system. Since the observed electrical signal is sufficiently large and the signal-to-noise ratio is reasonably good, down-scaling of sample dimensions to the nanometer-scale is feasible. 

\begin{acknowledgments}
We thank Stephen~Wu for assistance with ion milling. The work at Argonne was supported by the U.S. Department of Energy, Office of Science, Materials Science and Engineering Division. Lithography was carried out at the Center for Nanoscale Materials, an Office of Science user facility, which is supported by DOE, Office of Science, Basic Energy Science under Contract No. DE-AC02-06CH11357. The work at Colorado State University was supported by the U. S. Army Research Office (W911NF-14-1-0501), the U. S. National Science Foundation (ECCS-1231598), C-SPIN (one of the SRC STARnet Centers sponsored by MARCO and DARPA), and the U. S. Department of Energy (DE-SC0012670).
\end{acknowledgments}


\begin{thebibliography}{19}



%

	
\bibitem{STT}D.C.~Ralph and M.D.~Stiles, 
	J. Magn. Magn. Mater. \textbf{320}, 1190 (2008).

\bibitem{Mesoscale_mag}A.~Hoffmann and H.~Schulthei\ss,
	Curr. Opin. Solid State  Mater. Sci. (2014), doi:10.1016/j.cossms.2014.11.004.

\bibitem{Kajiwara}Y.~Kajiwara, K.~Harii, S.~Takahashi, J.~Ohe, K.~Uchida, M.~Mizuguchi~, H.~Umezawa, H.~Kawai, K.~Ando, K.~Takanashi, S.~Maekawa, and E.~Saitoh,
    Nature \textbf{464}, 262 (2010).

    
\bibitem{Jungfleisch}M.B.~Jungfleisch, A.V.~Chumak, V.I.~Vasyuchka, A.A.~Serga, B.~Obry, H.~Schultheiss, P.A.~Beck, A.D.~Karenowska, E.~Saitoh, and B.~Hillebrands,
        Appl. Phys. Lett. \textbf{99}, 182512 (2011). 
        

\bibitem{Jungfleisch_prop}M.B.~Jungfleisch, W.~Zhang, W.~Jiang, H.~Chang, J.~Sklenar, S.M.~Wu, J.E.~Pearson, A.~Bhattacharya, J.B.~Ketterson, M.~Wu, and A.~Hoffmann,
	J. Appl. Phys. \textbf{117}, 17D128 (2015).
	
	
	
	
\bibitem{Uchida}K.~Uchida, S.~Takahashi, K.~Harii, J.~Ieda, W.~Koshibae, K.~Ando, S.~Maekawa and E.~Saitoh,
	Nature \textbf{455}, 778 (2008).
	
\bibitem{Jungfleisch_SSE}M.B.~Jungfleisch, T.~An, K.~Ando, Y.~Kajiwara, K.~Uchida, V.I.~Vasyuchka, A.V.~Chumak, A.A.~Serga, E.~Saitoh and B.~Hillebrands,
	Appl. Phys. Lett. \textbf{102}, 062417 (2013).



\bibitem{Klein_PRL}A.~Hamadeh, O.~d'Allivy~Kelly, C.~Hahn, H.~Meley, R.~Bernard, A.H.~Molpeceres, V.V.~Naletov, M.~Viret, A.~Anane, V.~Cros, S.O.~Demokritov, J.L.~Prieto, M.~Mu\~noz, G.~de~Loubens, and O.~Klein,
	Phys. Rev. Lett. \textbf{113}, 197203 (2014).
	

\bibitem{Klein_APL}O.~d'Allivy~Kelly, A.~Anane, R.~Bernard, J.~Ben~Youssef, C.~Hahn, A.H.~Molpeceres, C.~Carr\'{e}t\'{e}ro, E.~Jacquet, C.~Deranlot, P.~Bortolotti, R.~Lebourgeois, J.-C.~Mage, G.~de~Loubens, O.~Klein, V.~Cros, and A.~Fert,
	Appl. Phys. Lett. \textbf{103}, 082408 (2013).
	

\bibitem{Hoffmann_book}M.~Wu and A.~Hoffmann, Recent advances in magnetic insulators - From spintronics to microwave applications, Solid State Physics \textbf{64}, (\textit{Academic Press}, 2013).


\bibitem{Melkov_book}A.G.~Gurevich and G.A.~Melkov, Magnetization oscillations and waves, \textit{CRC Press} (1996).




	
\bibitem{Pirro}P.~Pirro, T.~Br\"acher, A.V.~Chumak, B.~L\"agel, C.~Dubs, O.~Surzhenko, P.~G\"ornert, B.~Leven and B.~Hillebrands, 
	Appl. Phys. Lett. \textbf{104}, 012402 (2014).
	
	


\bibitem{Schneider}T.~Schneider, A.A.~Serga, A.V.~Chumak, C.W.~Sandweg, S.~Trudel, S.~Wolff, M.P.~Kostylev, V.S.~Tiberkevich, A.N.~Slavin, and B.~Hillebrands,
	Phys. Rev. Lett. \textbf{104}, 197203 (2010).

\bibitem{BEC}S.O.~Demokritov, V.E.~Demidov, O.~Dzyapko, G.A.~Melkov, A.A.~Serga, B.~Hillebrands, and A.N.~Slavin,
	Nature \textbf{443}, 430-433 (2006).

\bibitem{Kostylev_PRB}M.~Kostylev, V.E.~Demidov, U.-H.~Hansen, and S.O.~Demokritov,
	Phys. Rev. B \textbf{76}, 224414 (2007).
	
\bibitem{Houchen}H.~Chang, P.~Li, W.~Zhang, T.~Liu, A.~Hoffmann, L.~Deng, and M.~Wu, 
	IEEE Magn. Lett. \textbf{5}, 6700104 (2014).

	
\bibitem{Hoffmann}A.~Hoffmann,
	IEEE Trans. Magn. \textbf{49}, 5172 (2013).
	
	
\bibitem{Zhang_JAP}W.~Zhang, M.B.~Jungfleisch, W.~Jiang, J.~Sklenar, F.Y.~Fradin, J.E.~Pearson, J.B.~Ketterson, and A.~Hoffmann,
	J. Appl. Phys. \textbf{117}, 172610 (2015).
	
\bibitem{Zhang_AF}W.~Zhang, M.B.~Jungfleisch, W.~Jiang, J.E.~Pearson, A.~Hoffmann, F.~Freimuth, and Y.~Mokrousov,
	Phys. Rev. Lett. \textbf{113}, 196602 (2014).

\bibitem{Dyakonov}M.I.~D'yakonov and V.I.~Perel', 
	Sov. Phys. JETP Lett. \textbf{13}, 467 (1971).

\bibitem{Hirsch}J.E.~Hirsch, 
        Phys. Rev. Lett. \textbf{83}, 1834 (1999).





	
	

	
	
	
	

\bibitem{Liu_ST-FMR}L.~Liu, T.~Moriyama, D.C.~Ralph, and R.A.~Buhrman, 
	Phys. Rev. Lett. \textbf{106}, 036601 (2011).
	
\bibitem{Chiba}T.~Chiba, G.E.W.~Bauer, and S.~Takahashi,
	Phys. Rev. Applied \textbf{2}, 034003 (2014).
	
\bibitem{Chiba2}T.~Chiba, M.~Schreier, G.E.W.~Bauer, and S.~Takahashi,
	J. Appl. Phys. \textbf{117}, 17C715 (2015).	

\bibitem{Joe_PRL}J.~Sklenar, W.~Zhang, M.B.~Jungfleisch, W.~Jiang, H.~Chang, J.E.~Pearson, M.~Wu, J.B.~Ketterson, and A.~Hoffmann,
	ArXiv e-prints (2015), arXiv:1505.07791 [cond-mat.meshall].


\bibitem{Mellnik_ST-FMR}A.R.~Mellnik, J.S.~Lee, A.~Richardella, J.L.~Grab, P.J.~Mintun, M.H.~Fischer,	A.~Vaezi, A.~Manchon, E.-A.~Kim, N.~Samarth, and D.C.~Ralph, 
	Nature \textbf{511}, 449 (2014).

	




\bibitem{Demidov_Nat12}V.E.~Demidov, S.~Urazhdin, H.~Ulrichs, V.~Tiberkevich, A.~Slavin, D.~Baither, G.~Schmitz, and S.O.~Demokritov,
	Nat. Mater. \textbf{11}, 1028 (2012).


	
\bibitem{Slavin_bullet}A.~Slavin and V.~Tiberkevich,
	    Phys. Rev. Lett. \textbf{95}, 237201 (2005).
	   
\bibitem{Urazhdin}R.H.~Liu, W.L.~Lim, and S.~Urazhdin,
	Phys. Rev. Lett. \textbf{110}, 147601 (2013).
	

	





	

	
	
	


	
	



\bibitem{spin-diode}J.C.~Sankey, P.M.~Braganca, A.G.F.~Garcia, I.N.~Krivorotov, R.A.~Buhrman, and D.C.~Ralph, 
	Phys. Rev. Lett. \textbf{96}, 227601 (2006).

\bibitem{Nakayama}H.~Nakayama, M.~Althammer, Y.-T.~Chen, K.~Uchida, Y.~Kajiwara, D.~Kikuchi, T.~Ohtani, S.~Gepr\"ags, M.~Opel, S.~Takahashi, R.~Gross, G.E.W.~Bauer, S.T.B.~Goennenwein, and E.~Saitoh,
	Phys. Rev. Lett. \textbf{110}, 206601 (2013).




	
	
\bibitem{Schreier_STT}M.~Schreier, T.~Chiba, A.~Niedermayr, J.~Lotze, H.~Huebl, S.~Gepr\"ags, S.~Takahashi, G.E.W.~Bauer, R.~Gross, and S.T.B.~Goennenwein, 
	ArXiv e-prints (2014), arXiv:1412.7460 [cond-mat.meshall].




	

	    


\bibitem{Tserkovnyak}Y.~Tserkovnyak, A.~Brataas and G.E.W.~Bauer, 
    Phys. Rev. Lett. \textbf{88}, 117601 (2002).
	
\bibitem{Azevedo_ISHE}A.~Azevedo, L.H.~Vilela ~Le\~{a}o, R.L.~Rodr\'{i}guez-Su\'{a}rez, A.B.~Oliveira, and S.M.~Rezende,
	J. Appl. Phys. \textbf{97}, 10C715 (2005).
	

	
\bibitem{Mosendz}O.~Mosendz, J.E.~Pearson, F.Y.~Fradin, G.E.W.~Bauer, S.D.~Bader, and A.~Hoffmann,
    Phys. Rev. Lett. \textbf{104}, 046601 (2010).

	
	



\bibitem{BLS1}B.~Hillebrands,
	Brillouin light scattering spectroscopy, in Novel techniques for characterising magnetic materials, edited by Y. Zhu, \textit{Springer} (2005).




\bibitem{mumax3}A.~Vansteenkiste, J.~Leliaert, M.~Dvornik, M.~Helsen, F.~Garcia-Sanchez, and B.~Van~Waeyenberge,
	AIP Advances \textbf{4}, 107133 (2014). 
	

	

\bibitem{Zhang_APL13}W.~Zhang, V.~Vlaminck, J.E.~Pearson, R.~Divan, S.D.~Bader, and A.~Hoffmann, 
	Appl. Phys. Lett. \textbf{103}, 242414 (2013).

	






\bibitem{edge_mode}J.~Jorzick,S.O.~Demokritov, B.~Hillebrands, M.~Bailleul, C.~Fermon, K.Y.~Guslienko, A.N.~Slavin, D.V.~Berkov, and N.L.~Gorn,
	Phys. Rev. Lett. \textbf{88}, 047204 (2002).
	
\bibitem{Neusser}S.~Neusser and D.~Grundler,
	Advanced Materials \textbf{21}, 2927 (2009).



\end{thebibliography}
\end{document}